\documentclass[%
reprint,
 amsmath,amssymb,
 aps,
]{revtex4-2}

\usepackage{color}
\usepackage{graphicx}
\usepackage{dcolumn}
\usepackage{braket}
\usepackage{bm}

\usepackage{amsthm}

\newcommand{\bs}[1]{\boldsymbol{\mathbf{#1}}}
\newcommand{\shift}[1]{\mbox{#1-\ss}}

\begin{document}

\title{
No eigenstate of the critical transverse-field Ising chain satisfies the area law
}
\author{Saverio Bocini}
\affiliation{%
 Universit\'e Paris-Saclay, CNRS, LPTMS, 91405, Orsay, France
}%
\author{Maurizio Fagotti}
 \email{maurizio.fagotti@universite-paris-saclay.fr}
\affiliation{%
 Universit\'e Paris-Saclay, CNRS, LPTMS, 91405, Orsay, France
}%

\begin{abstract}
We argue that, in a basis common to all one-site shift invariant  conserved charges, there is no eigenstate of a noninteracting local spin-$\frac{1}{2}$ chain Hamiltonian that satisfies the area law if the ground state has half-integer central charge.
That is to say, in those models all (quasi)local one-site shift invariant conserved operators are gapless. From the standpoint of  bipartite entanglement properties, we show indeed that
there are three distinct one-site shift invariant noninteracting models, two of which are  equivalent  to the XX model (for one of them the transformation breaks one-site shift invariance) and the other to the critical Ising model. The former class has two locally distinct one-site shift invariant excited states satisfying the area law; the latter two classes have none. 
\end{abstract}

\maketitle

The ground states of shift-invariant spin-chain systems with gapped local Hamiltonians are low entangled: the entropy of a block of spins has only a subleading dependence on the block's length~\cite{Hastings2007An}. This is known as ``area law'' and applies to systems in higher dimensions as well~\cite{Eisert2010Colloquium}. In 1D the area law generally breaks down at quantum phase transitions, where the entropy of a spin block can develop a logarithmic dependence on its length~\cite{Calabrese2009Entanglement}. In contrast, excited states are expected to follow a volume law~\cite{Hayden2006Aspects}: the entropy of a block of spins is proportional to the block's length. Exceptions to this rule (nearly) define the so-called quantum many-body scars~\cite{Serbyn2021Quantum}, which are excited states with (generally) anomalously low bipartite entanglement. Integrable systems are quite exceptional in this respect, as they exhibit infinitely many excited states with sub-extensive entropies and  energies that are extensively larger than the ground state one~\cite{Alba2009Entanglement,Beugeling2015Global}. This is a direct consequence of the existence of infinitely many local conservation laws, indeed the ground state of any conserved charge with fast enough decaying interactions is an eigenstate of the Hamiltonian with sub-extensive bipartite entanglement. 
Both the typical properties of excited states~\cite{Alba2009Entanglement,Bhattacharya2013,Wong2013,Palmai2014Excited, Molter2014Bound,Ares2015,Beugeling2015Global,Lai2015Entanglement,Alba2015Eigenstate,Keating2015,Lev2017Entanglement,Lev2017Entanglement1,Sanjay2018,Lev2018Volume,Mestyan2018Renyi,Moessner.Haque2018,Hackl2019Average,Murthy2019Structure,LeBlond2019,Murciano2019,Levine2019,Lu2019Renyi,Moitra2020,LydZba2020Eigenstate,Angel-Ramelli2021,Miao2021Eigenstate,Miao2022Eigenstate,Haque2022Entanglement,Bianchi2022Volume,Buividovich2022unpublished,Yu2023Free} and the critical properties of low-entangled ones~\cite{Alcaraz2011Entanglement,Berganza2012Entanglement, Storms2014Entanglement,Herwerth2015Excited,Jafarizadeh2019Bipartite, Zhang2021Corrections, Zhang2022Entanglement} have attracted some attention.
How many locally different excited states satisfy however the area law? This is a tricky, somewhat ill-defined question that has been either overlooked or addressed just incidentally and/or carelessly. 
On one hand, the answer could depend on the basis chosen to diagonalise the Hamiltonian and degeneracies could also be sensitive to the system's size. On the other hand, the choice of the basis can be physical, for example, in investigations into the stationary behaviour of observables when the system is prepared in some nonequilibrium initial state. 
That was our very motivation for starting this investigation. In contrast to the ordinary behaviour reported in Refs~\cite{Maric2022Universality,Maric2023Universality},  we have indeed observed an unusually slow decay of connected correlations at late times after a quench between one-site shift invariant (``\shift{1}'' in the following) critical Hamiltonians in the Ising universality class. What is then so special in the excited states of the critical Ising model?
In \shift{1} noninteracting spin chains that are mapped to free fermions by a Jordan-Wigner transformation, such as the quantum Ising model, a relevant basis of excited states consists of \shift{1} Slater determinants for the Jordan-Wigner fermions.  
In any such excited state, the bipartite entanglement entropies can be easily computed numerically, and the entropies of large subsystems can be predicted on the basis of the asymptotic behaviours of determinants of Toeplitz and block-Toeplitz matrices~\cite{Its2009The}, which have been thoroughly investigated (see, e.g.,  Refs~\cite{Deift2011Asymptotics,Basor2017Asymptotics}). This opportunity is definitely rare and has already been  exploited to quantify the picture summarised above~\cite{Alba2009Entanglement,Ares2014,Jafarizadeh2019Bipartite}. 
In particular, it was shown that there are infinitely many excited states that can be described by zero-temperature conformal field theories (CFTs) with half-integer or integer central charges. More vague are the statements about excited states satisfying the area law. 
It looks like there is a common belief that one should expect infinitely many states of that kind. We point out here that the picture is quite different. Specifically, only a finite number of locally different excited states seem to satisfy the area law, and there are notable cases, such as the critical Ising model, where there are none. 
We argue that this no-go theorem applies to every noninteracting \shift{1} local Hamiltonian with a critical ground state described by a CFT with half-integer central charge. 
We show it in two steps. First, we explicitly study the XX model~(XX)
\begin{equation}\label{eq:XX}
\bs H_{\textsc{xx}}=J\sum\nolimits_\ell\bs\sigma_\ell^x\bs\sigma_{\ell+1}^x+\bs\sigma_\ell^y\bs\sigma_{\ell+1}^y\, ,
\end{equation}
the critical Ising model (CI)
\begin{equation}\label{eq:CI}
\bs H_{\textsc{ci}}=-J\sum\nolimits_\ell\bs\sigma_\ell^x\bs\sigma_{\ell+1}^x+\bs\sigma_\ell^z\, ,
\end{equation}
and the strongly anisotropic XY model (X-X)
\begin{equation}\label{eq:X-X}
\bs H_{\textsc{x-x}}=J\sum\nolimits_\ell\bs\sigma_\ell^x\bs\sigma_{\ell+1}^x-\bs\sigma_\ell^y\bs\sigma_{\ell+1}^y\, .
\end{equation}
We aver that, among them, only the XX model has eigenstates satisfying the area law (in the \shift{1} basis); and no excited state  is described by a CFT with half-integer central charge.
We then prove that every \shift{1} noninteracting Hamiltonian can be mapped into a \shift{1} conservation law of XX, CI, or \mbox{X-X}   by a discrete and/or a continuous local unitary transformation. Such transformations do not affect the asymptotic dependency of the bipartite entropies on the subsystem's length, hence the generality of our findings.

\paragraph*{Noninteracting spin chains.}
We consider a spin-$\frac{1}{2}$ chain with $L$ sites described by a generic \shift{1} Hamiltonian (with periodic boundary conditions) 
that can be mapped to free fermions by a Jordan-Wigner transformation $\bs a_{2\ell-1}=\prod_{j<\ell}\bs\sigma_j^z \bs\sigma_\ell^x$, $\bs a_{2\ell}=\prod_{j<\ell}\bs\sigma_j^z \bs\sigma_\ell^y$, where $\bs a_j$ are Majorana fermions ($\{\bs a_i,\bs a_j\}=2\delta_{i j}$) which in the following will be imagined as sitting on pseudo-sites ($j$ in $\bs a_j)$.  
The excited states, $\ket{\{p\}}$, split in two sectors, usually called Ramond (+) and Neveu-Schwarz (-), differing only in the quantization conditions satisfied by the momenta $p$ of the excitations: $e^{iL p}=1$ in Ramond and $e^{iL p}=-1$ in Neveu-Schwarz. 
An excited state is in Ramond if $\prod_{j=1}^L\bs\sigma_\ell^z\ket{\{p\}}=-\ket{\{p\}}$ and in  Neveu-Schwarz  if $\prod_{j=1}^L\bs\sigma_\ell^z\ket{\{p\}}=\ket{\{p\}}$; 
in the fermionic picture it is an eigenstate of a quadratic form of fermions $\bs H^\pm=\frac{1}{4}\sum_{\ell,n}\bs a_\ell\mathcal H^\pm_{\ell n}\bs a_n$ with periodic ($+$) or anti-periodic ($-$) boundary conditions (i.e., $\mathcal H^+$ is block circulant and $\mathcal H^-$ block anti-circulant), respectively.
As such, each excited state is fully characterised by the fermionic two-point correlations, which are customarily organised in a matrix $\Gamma_{\{p\}}=\mathrm I-\braket{\{p\}|\bs a\otimes \bs a|\{p\}}$, known as correlation matrix. 
Translational invariance makes it convenient to work in the Fourier space.  
We refer the reader to Ref.~\cite{Fagotti2016Charges} for a review of some free-fermion techniques; here we only list some useful  well-known results.  
The $2$-by-$2$ block-Fourier transform (aka symbol) $\hat \Gamma^{(2)}(k)$ of a correlation matrix $\Gamma$ is $2\pi$-periodic and satisfies $\hat \Gamma^{(2)}(k)=[\hat \Gamma^{(2)}(k)]^\dag=-[\hat \Gamma^{(2)}(-k)]^t$ (by the fermionic algebra)~\footnote{We are defining a block-Fourier transform rather than a simple Fourier transform because there are two Majorana fermions per site.}. Restricting to excited states, $\hat \Gamma^{(2)}_{\{p\}}(k)$
can be written in terms of the block-Fourier transform $\hat{\mathcal H}^{(2)}(k)$ of $\mathcal H^\pm$ as follows. First, $\hat \Gamma^{(2)}_{\{p\}}(k)$ commutes with the symbol of the Hamiltonian $\hat{\mathcal H}^{(2)}(k)$ and, for given $k$, has eigenvalues $\pm 1$;
each choice of eigenvalues corresponds to a different excited state, and, in particular, for the ground state we have $
\hat \Gamma^{(2)}_{\textsc{gs}}(k)=-\mathrm{sgn}(\hat{\mathcal H}^{(2)}(k))
$.
Second, in the basis diagonalising $\hat{\mathcal H}^{(2)}(k)$ in such a way that the first eigenvalue is identified with the excitation energy $\varepsilon(k)$ (and the second is, in turn, $-\varepsilon(-k)$), adding an excitation with momentum $\bar p$ to $\ket{\{p\}}$ ($\bar p\notin \{p\}$) has a double effect~\footnote{Note that only an even number of excitations preserves the sector.}: $\Sigma_{\{p\}\cup \{\bar p\}}(\pm \bar p)= \mp \sigma^z \Sigma_{\{p\}}(\pm \bar p)$, where $\Sigma_{\{p\}\cup \{\bar p\}}(\pm \bar p)$ is the diagonal matrix equivalent to the symbol of the correlation matrix (see also Ref.~\cite{Alba2009Entanglement}).

\paragraph*{Entanglement entropies.}
The entanglement properties of a bipartition $A\cup \overline A$ in a pure state can be measured by the von Neumann entropy $S_1(A)=-\mathrm{tr}[\rho_A\log \rho_A]$~\cite{Bennett1996Concentrating}, or, more generally, by the R\'enyi entropies $S_\alpha(A)=\log\mathrm{tr}[\rho_A^\alpha]/(1-\alpha)$~\cite{Renyi1970Probability}. 
Let $\bs O_A$ be an observable acting nontrivially only in a connected subsystem $A$. By the Wick's theorem, its expectation value in an excited state $\ket{\{p\}}$ can be expressed in terms of the submatrix $\Gamma_{\{p\},A}$ of $\Gamma_{\{p\}}$ corresponding to restricting the indices to the sites in $A$.  
$A$'s correlation matrix $\Gamma_{\{p\},A}$ is in turn block-Toeplitz  with symbol $\hat\Gamma_{\{p\}}^{(2)}(k)$. From now on we use the notation $\mathrm T_{N}^{(n)}$ to indicate $(nN)$-by-$(nN)$ block-Toeplitz matrices with $n$-by-$n$ blocks.
Several representations of the R\'enyi entropies of a spin block $A$ in terms of $\Gamma_{\{p\},A}$ are known. Here we use a formula that can be argued from the Euclidean approach to the R\'enyi entropies in free quantum field theories~\cite{Casini2009Entanglement,Maric2023Universality}
\begin{equation}\label{eq:Salpharep}
S_\alpha(A)=\sum_{j=0}^{\alpha-1}\frac{\log\det\bigl|\varrho_{|A|}^{(2)}(\frac{\pi(j+\frac{1}{2})}{\alpha})\bigr|}{2(1-\alpha)}\, ,
\end{equation}
where $\varrho_{|A|}^{(2)}(\phi)$ has the 2-by-2 symbol
\begin{equation}
\hat \varrho^{(2)}(k;\phi)=\sin\phi\mathrm I_2-i \cos\phi\hat \Gamma_{\{p\}}^{(2)}(k)\, .
\end{equation} 
This allows us to carry out a qualitative analysis of bipartite entanglement without specifying the R\'enyi index~$\alpha$.  
Before starting on it, we warn the reader that the entanglement properties of excited states could become basis dependent especially in superintegrable (aka non-Abelian integrable) systems, such as the quantum XY model in zero field~\cite{Fagotti2014On}, or in integrable systems with Hilbert-space fragmentation~\cite{Moudgalya2022Hilbert}, such as the dual folded XXZ model~\cite{Zadnik2021The}, in which there are exponentially large degenerate sectors spanned by low-entangled states. 
In our case degeneracies can be originated by  the symmetries of the dispersion relation or by accidental equivalences of multi-particle energies.  
Generally, both kinds of degeneracies can be lifted by adding a \shift{1} conservation law to the Hamiltonian and/or by changing the chain's length. This selects a natural \shift{1} basis of excited states. 

Our first step will be an analytical study, which is mainly a reorganisation of results that are already known or that can be readily derived from what is  known. 
\paragraph*{XX model.}
The XX model has the $U(1)$ symmetry of rotations about $z$, i.e., the Hamiltonian commutes with $\bs S^z=\frac{1}{2}\sum_\ell\bs\sigma_\ell^z$. This is enough to conclude that there are at least two locally different excited states satisfying the area law: $\ket{\cdots \uparrow\uparrow\cdots}$ and $\ket{\cdots \downarrow\downarrow\cdots}$. Our  goal in this case is to establish that any other state of that kind is locally equivalent to one or to the other.  Let us start by writing the symbol of the correlation matrix in $\ket{\{p\}}$:
\begin{equation}
\hat \Gamma_{\{p\}}^{(2)}(k)=\tfrac{\hat \Sigma_{\{p\}}^{(1)}(k)-\hat \Sigma^{(1)}_{\{p\}}(-k)}{2}\mathrm I_2+\tfrac{\hat \Sigma^{(1)}_{\{p\}}(k)+\hat \Sigma^{(1)}_{\{p\}}(-k)}{2}\sigma^y\, ,
\end{equation}
where $\hat \Sigma^{(1)}_{\{p\}}(k)\in\{-1,1\}$ and, in particular, $\hat \Sigma^{(1)}_{\emptyset}(k)=-\mathrm{sgn}(\cos k)$.
The simple structure of the symbol is translated into a simple structure of the correlation matrix, which allows one to express the entropies as determinants of $|A|$-by-$|A|$ Toeplitz matrices: 
\begin{equation}\label{eq:logdetXX}
\det\varrho_A(\phi)=\Bigl|\det\bigl|\sin\phi\mathrm I_{|A|}+i \cos\phi\Sigma_{|A|}^{(1)}\bigr|\Bigr|^2\, ,
\end{equation}
where $\Sigma_{|A|}^{(1)}$ has symbol $\hat \Sigma^{(1)}_{\{p\}}(k)$. 
In the thermodynamic limit we can  apply the result proved by Szeg\"o in Ref.~\cite{Szego1920Beitrage}, and obtain the asymptotic behaviour as $|A|\rightarrow\infty$ 
\begin{equation}\label{eq:ext}
\frac{\log\det\varrho_A(\phi)}{|A|}\rightarrow \int_{-\pi}^\pi\frac{\mathrm d k}{2\pi}\log|\sin^2\phi+\cos^2\phi [\hat \Sigma^{(1)}(k)]^2|\, .
\end{equation}
Here we dropped the label $\{p\}$ because now  $\hat \Sigma^{(1)}(k)$ is the symbol of a macrostate, representing infinitely many locally equivalent excited states, and can assume any value in $[-1,1]$.  
The excited states with minimal entropy are sub-extensive and, in turn, their $\hat \Sigma^{(1)}(k)$ should make the right hand side of \eqref{eq:ext} vanish, that is to say,
$
\hat\Sigma^{(1)}(k)\in\{-1,1\}
$.  
All sequences of excited states with sub-extensive entropy  are therefore characterised by a piece-wise constant symbol $\hat \Sigma^{(1)}(k)$ with a given number $n$ of discontinuities. If the latter is finite, the discontinuities are simple examples of Fisher-Hartwig singularities~\cite{Basor1983Toeplitz,Deift2011Asymptotics}. Specifically, each discontinuity gives the additive asymptotic contribution $-(\frac{\phi}{\pi}-\frac{1}{2})^2\log|A|$ to $\log \det|\sin\phi\mathrm I_{|A|}+i \cos\phi\Sigma_{|A|}^{(1)}|$ in \eqref{eq:logdetXX}.
The corresponding states behave as ground states of CFTs with integer central charges $\frac{n}{2}$. Their symbol can indeed be reinterpreted as the sign of the symbol of a quasilocal quadratic form of fermions and vice versa. 
The area law can be satisfied only if there are no discontinuities at all, i.e., $\hat \Sigma^{(1)}(k)=\pm 1$. These are 
the symbols of the states that are locally equivalent to $\ket{\cdots \uparrow\uparrow\cdots}$ and $\ket{\cdots \downarrow\downarrow\cdots}$. 
Consequently, the ground state of any gapped conservation law with exponentially decaying interactions is either $\ket{\cdots \uparrow\uparrow\cdots}$ or $\ket{\cdots \downarrow\downarrow\cdots}$.
It can also be argued --- \footnote{See the Supplemental Material.}-D --- that there should not be exceptions in the presence of infinitely many discontinuities in the symbol, as the bipartite entanglement tends to increase with their number (such states can also be interpreted as ground states of long-range Hamiltonians~\cite{Gori2015Explicit}).

\paragraph*{Critical Ising model.}
The critical Ising model is special as $\mathcal H^\pm$ is not only antisymmetric block-\mbox{(anti)circulant}, but it is  antisymmetric \mbox{(anti)circulant} (antisymmetry manifests the fermionic algebra). 
This property can be almost directly translated to the correlation matrices of the excited states. There is only a subtlety due to the 2-fold degeneracy of the eigenstates in the Ramond sector, which break the invariance under one pseudo-shift --- see~\cite{Note3}-A. Since our search for states satisfying the area law is not affected by this issue, when needed, we will consider the incoherent superposition of the two degenerate states, which is described by a circulant matrix $\Gamma^{(1)}$ with symbol $\hat \Gamma^{(1)}_{\{p\}}(k)\in \{-1,1\}$ for $k\neq 0,\pi$ and $\hat \Gamma^{(1)}(0)=\hat \Gamma^{(1)}(\pi)=0$. 
We then have
\begin{equation}\label{eq:varrhoCI}
\det\varrho_A(\phi)=\det\bigl|\sin\phi\mathrm I_{2|A|}+i \cos\phi\Gamma_{2|A|}^{(1)}\bigr|\, .
\end{equation}
In the thermodynamic limit the Szeg\"o lemma gives
\begin{equation}
\frac{\log\det\varrho_A(\phi)}{|A|}\rightarrow \int_{-\pi}^\pi\frac{\mathrm d k}{2\pi}\log|\sin^2\phi+\cos^2\phi [\hat \Gamma^{(1)}(k)]^2|,
\end{equation}
which matches the expression for the XX model provided that  $\Sigma^{(1)}(k)$ is replaced by $\Gamma^{(1)}(k)$. 
The qualitative analysis is therefore almost identical. The unique but important deviation comes from the fact that, while $\Sigma^{(1)}(k)$ could be any function with image in $[-1,1]$, $\Gamma^{(1)}(k)$ is forced to be odd by the fermionic algebra. Consequently, the smallest number of discontinuities in the symbol is $2$ (one at $k=0$ and one at $k=\pi$) and corresponds to the ground state or to the state with maximal energy. These states break the area law with a logarithmic term corresponding to a CFT with central charge $\frac{1}{2}$. Contrary to the XX model, there is no excited state satisfying the area law that can be interpreted as the ground state of a quasilocal Hamiltonian, that is to say, every quasilocal conservation law of the critical Ising model is gapless. 
As also observed in Ref.~\cite{Jafarizadeh2019Bipartite}, we find that the central charges of the CFTs describing the low-entangled excited states of CI can be both half-integer and (nonzero) integer. 
\paragraph*{X-X model.}
The X-X model, described by Hamiltonian~\eqref{eq:X-X}, is equivalent to the XX model, indeed $\bs H_{\textsc{X-X}}=\bs\Pi^{zy} \bs H_{\textsc{XX}} \bs\Pi^{zy}$, where $\bs\Pi^{zy}=\prod_{j}\bs\sigma_{2j-1}^z\bs\sigma_{2j-1}^y$. The similarity transformation, however, breaks one-site shift invariance, hence only the \shift{1} charges of the XX model that are even under $\bs\Pi^{zy}$ are mapped to \shift{1} charges of the X-X model. The remaining \shift{1} charges of X-X correspond to conservation laws of $XX$ that break one-site shift invariance into a two-site one, whose existence was established in Ref.~\cite{Fagotti2014On}. Thus, we are  forced to treat the X-X model  independently of the XX one. 

The symbol of the correlation matrix in $\ket{\{p\}}$ reads
\begin{equation}
\Gamma_{\{p\}}^{(2)}(k)=\tfrac{\hat \omega_{+,\{p\}}(k)+\hat \omega_{-,\{p\}}(k)}{2}\mathrm I+\tfrac{\hat \omega_{+,\{p\}}(k)-\hat \omega_{-,\{p\}}(k)}{2}\sigma^x\, ,
\end{equation}
where $\hat \omega_{s,\{p\}}(k)\in \{-1,1\}$ are odd in $k$. Just as we did in CI, we are overlooking  the zero energy modes $k=0,\pi$, which would give a contribution $(s_0 \delta_{k 0}+s_\pi \delta_{k \pi})\sigma^y$ --- see~\cite{Note3}-A. 
We can again effectively decompose the block-Toeplitz matrix so as to express the entropies in terms of determinants of Toeplitz matrices
\begin{equation}
\det\varrho_A(\phi)= \prod_{s=\pm 1}\det\bigl|\sin\phi\mathrm I_{|A|}+i \cos\phi\omega_{s,|A|}^{(1)}\bigr|\, .
\end{equation}
The right hand side is the product of two determinants of the Ising kind~\eqref{eq:varrhoCI}, therefore the entanglement entropies are  the sum of the entropies of two excited states of the Ising model. The absence of a state satisfying the area law in the critical Ising model implies in turn the same in the X-X model. 
Remarkably, also here the central charges of the CFTs describing the low-entangled excited states can be both half-integer and integer.

\paragraph*{Numerical analysis.}
\begin{figure}
    \centering
    \includegraphics[width=\linewidth]{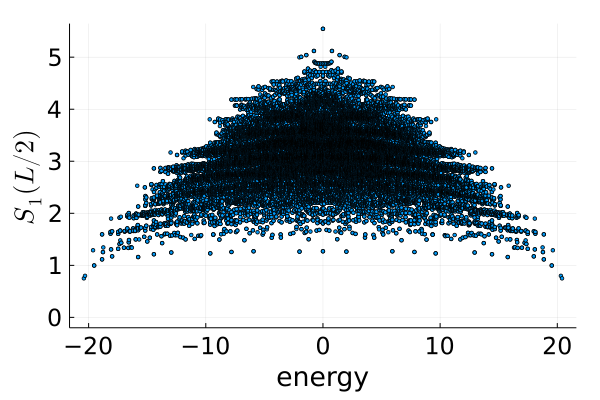}
    \caption{Entropy of all the \shift{1} eigenstates of CI as a function of their energy for $L=16$.
    }
    \label{fig:scar_plot}
\end{figure}
Although persuasive, our analytical hints are based on the assumption that the local properties of any excited state can be described with a symbol that does not depend on the subsystem's length; but this is too restrictive. 
We can indeed easily imagine exotic sequences of excited states that do not fit in the classification above: e.g., for any given $|A|$, we can find excited states in which the density of excitations is $\sim \frac{\log|A|}{|A|}$, contaminating the family of states that break the area law logarithmically by states that satisfy the volume law. 
In order to rule out unconventional states satisfying the area law, we resort to a numerical analysis. 

We start considering finite chains and evaluate the half-chain von Neumann entropy of all excited states. 
Those satisfying the area law correspond to minima of such an entropy landscape that remain bounded even increasing the chain's length $L$. For small $L$, we can recognise the global minima by a brute-force search. This is shown in  Fig.~\ref{fig:scar_plot} for CI, which suggests that the ground state and the maximal energy state have the lowest bipartite entanglement and points, in turn, at the absence of excited states satisfying the area law. A real quantitative difference between states satisfying or breaking the area law appears however only in much larger chains.  
The number of eigenstates, on the other hand, diverges exponentially with $L$, thus computing the entropy of all eigenstates becomes soon unfeasible. 

\begin{figure}
    \centering
    \includegraphics[width=\linewidth]{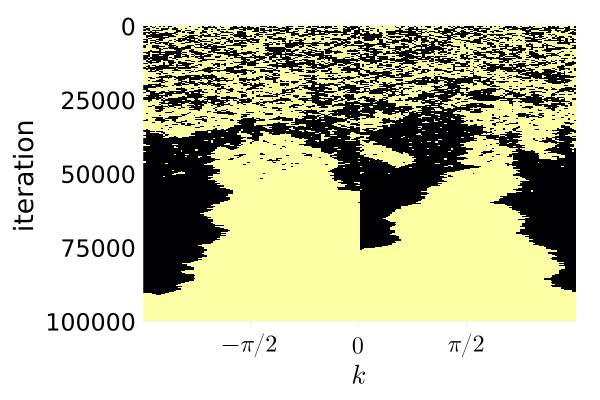}
    \caption{A single run of simulated annealing  converging to the ground state of CI in the Neveu-Schwarz sector with $L=200$. 
    In yellow (black) the particle (hole) momenta.
    }
    \label{fig:percolation_plot}
\end{figure}
We circumvent this problem by means of simulated annealing:
We initialise the system in a random eigenstate $\ket{\{p\}}$ generated from a uniform distribution. At each step we make a particle-hole transformation at a random momentum and accept the new excited state $\ket{\{p'\}}$ with probability $\min\{\exp(\frac{S_1(\frac{L}{2};\{p\})-S_1(\frac{L}{2};\{p'\})}{\mathfrak s}, 1\}$ in analogy with the Metropolis–Hastings algorithm.
Here $\mathfrak s$ is a parameter that depends on the iteration and is chosen so as to approach $0$ at the end of the simulation~(more details about the algorithm are reported in \cite{Note3}-B). 
Fig.~\ref{fig:percolation_plot} shows a run of it in CI. In a reasonable number of iterations the excited state becomes of the form that we investigated analytically (i.e., the number of discontinuities in its symbol becomes independent of $L$) and finally the algorithm selects the ground state or the maximal energy state.  Similar analyses for XX and X-X have confirmed the pictures drawn before and are reported in~\cite{Note3}-C. 

\paragraph*{A lower bound.} The last proof we give is based on a conjecture. In~\cite{Note3}-D we provide evidence that the half-chain von-Neumann entropy of an XX excited state for which $\hat \Sigma^{(1)}_{\{p\}}(k)$ has $n$ changes of signs satisfies
\begin{equation}\label{eq:conjecture}
S_1(A)\geq (\tfrac{\log 3}{2}-\tfrac{\log 2}{3}) n\, ,
\end{equation}
where $|A|=\frac{L}{2}$.
For CI, if we consider the discontinuities of the scalar symbol, an analogous expression applies with an overall factor $1/2$.  The area law can only hold if $n$ is finite. Since two locally different states differ in an extensive number of excitations, such a state would be always locally equivalent to one of the states that we addressed in the thermodynamic limit with a finite number of discontinuities smaller than or equal to~$n$. Thus, \eqref{eq:conjecture} implies our conclusions.  

\paragraph*{Equivalence of \shift{1} quasilocal models.}
Ref.~\cite{Bravyi2006Lieb} showed 
that a local operator that time evolves for a finite time under a local Hamiltonian can still be approximated by another local operator with an error (in operator norm) that decreases exponentially with its range. 
Such transformations are used to define topological phases of matter~\cite{Zeng2019book} and will be denoted by $\mathfrak{L}$. They are relevant to us because they preserve the asymptotic dependency of the bipartite entropies on the subsystem's length. 
To that aim, however, one can also consider discrete transformations, denoted by $\mathfrak{D}$, that do not have a local generator but act as $\mathfrak{L}$ on local operators. An example is a shift by a finite number of sites. Since we are studying the entropies of connected blocks, we can overlook the difference between spins and fermions~\cite{Vidal2003Entanglement,Fagotti2010disjoint} and  preserve locality in whichever of the two spaces. This allows us to include also transformations such as a shift by one pseudo-site. In~\cite{Note3}-E we show 
\begin{equation}
\hat {\mathcal H}^{(2)}(k)\overset{\mathfrak{L}}\mapsto\begin{cases}
e^{i n k \sigma^z}\sigma^y &\overset{\mathfrak{D}}{\mapsto}\ \sigma^y\\
\sin ke^{i n k \sigma^z}\sigma^x&\overset{\mathfrak{D}}{\mapsto}\ \sin k\sigma^x\\
\cos \tfrac{k}{2}  e^{i (n-\frac{1}{2}) k \sigma^z}\sigma^y&\overset{\mathfrak{D}}{\mapsto}\ \sin \tfrac{k}{2} e^{i \frac{k }{2} \sigma^z}\sigma^x \\
\sin \tfrac{k}{2} e^{i (n-\frac{1}{2}) k \sigma^z}\sigma^x &\overset{\mathfrak{D}}{\mapsto}\ \sin \tfrac{k}{2} e^{i \frac{k }{2} \sigma^z}\sigma^x 
\end{cases}
\end{equation}
where all symbols are defined up to multiplication by a smooth even function. 
In other words, we recognise four families of topologically different models. After additional discrete transformations,  the first family is mapped to XX, the second one to X-X, and the last two of them to CI.
Since only in XX there are excited states satisfying the area law but there are no quasilocal conserved operators whose ground state is described by a half-integer central charge, we can conclude that, quite generally, there are no excited states satisfying the area law when the ground state of a \shift{1} noninteracting Hamiltonian is described by a CFT with half-integer central charge.
\paragraph*{Discussion.} 
We have reported a no-go theorem connecting the  central charge of the conformal field theory describing the low-energy properties of a noninteracting spin chain with the existence of excited states satisfying the area law.  
Concerning the generality of our findings, there are several open questions. Can the assumption of one-site shift invariance be partially relaxed? Does something similar apply also in the presence of interactions? 
In the thermodynamic limit any linear combination of excited states in a collapsing energy shell is conserved to all intents and purposes; does the result hold true also in such quasi-stationary states? 

\begin{acknowledgments}
We thank Vanja Mari\'c and Mikhail Zvonarev for discussions. 
This work was supported by the European Research Council under the Starting Grant No. 805252 LoCoMacro.
\end{acknowledgments}

\bibliography{references}

\section*{Supplemental Material}
\twocolumngrid

\subsection{On the zero energy modes}\label{SM:zero}
We remind the reader that the symbol of the correlation matrix of an excited state
commutes with the symbol of the Hamiltonian (otherwise the correlation matrix would have a nontrivial time evolution)~\cite{Alba2009Entanglement, Fagotti2016Charges}. 
In addition, for the state to be pure, its symbol should have eigenvalues $\pm 1$. 
These two conditions, together with the standard properties that a symbol is required to satisfy ($\hat T(k)=\hat T^\dag(k)=-\hat T^{t}(-k)$), completely characterise the symbols of the excited states. 

Remarkably, excited states with the same symmetries of the Hamiltonian do not necessarily exist. In particular, the smallest symbol representing an excited state could have a different size with respect to the smallest symbol representing the Hamiltonian.
This is what happens in the critical Ising model (CI). 
The scalar symbol of the CI's Hamiltonian, indeed, reads
$$
\hat{\mathcal H}^{(1)}(k)=\sin k\, .
$$
This clearly commutes with any other scalar function, but there is no scalar symbol that meets the conditions above in the Ramond sector: whatever function we choose, it will always have two eigenvalues equal to zero (associated with $k=0$ and $k=\pi$). Hence, the state is not pure. Pure excited states in the Ramond sector have a symbol that is at least 2-by-2. Specifically, it has the form
$$
\hat\Gamma^{(2)}(k)=\tfrac{\omega(\frac{k}{2})+\omega(\frac{k}{2}+\pi)}{2}\mathrm I_2+\tfrac{\omega(\frac{k}{2})-\omega(\frac{k}{2}+\pi)}{2}\sigma^x e^{-i\frac{k}{2}\sigma^z}+s \delta_{k 0}\sigma^y
$$
where $\omega(k),s\in \{-1,1\}$ and $\omega(-k)=-\omega(k)$. This symbol would have had also a scalar representation if $s$ could be set to zero. 
Instead, the excited states break the pseudo-shift invariance of the Ramond Hamiltonian  into \shift{1} invariance. 
In practice, the effect of $s$ in the correlation matrix is to introduce the following corrective term
$$
\frac{s}{L} \mathrm J_L\otimes \sigma^y
$$
where $\mathrm J_L$ is the $L$-by-$L$ matrix of ones. In the thermodynamic limit this term is irrelevant because it gives a correction $O(\frac{|A|}{L})$. The correction to the half-chain von-Neumann entropy is bounded as well by $\log 2$ because the mixed state ($s=0$) consists of the incoherent superposition of just two states ($s=\pm 1$). Nevertheless, in our numerical investigations we use the exact symbol. 

An analogous situation applies to the symbols of the excited states in the X-X model, where we have
\begin{multline}
\Gamma_{\{p\}}^{(2)}(k)=\tfrac{\hat \omega_{+,\{p\}}(k)+\hat \omega_{-,\{p\}}(k)}{2}\mathrm I+\tfrac{\hat \omega_{+,\{p\}}(k)-\hat \omega_{-,\{p\}}(k)}{2}\sigma^x+\\
(s_0 \delta_{k 0}+s_\pi \delta_{k \pi})\sigma^y\, ,
\end{multline}
with $\hat \omega_{s,\{p\}}(k)\in \{-1,1\}$  odd functions of $k$ and $s_0,s_\pi\in \{-1,1\}$ parametrize the ambiguity in the zero energy modes $k=0,\pi$. 

\subsection{On the numerical simulations}\label{SM:num}
We have numerically searched for the state with minimal entropy  in each fermionic sector (Neveu-Schwarz and Ramond) separately.
Those eigenstates are shared by the original spin Hamiltonian if they have the right parity of the number of excitations, i.e., $\prod_j \bs\sigma_j^z \ket{\{p\}}=-\ket{\{p\}}$ in Ramond and $\prod_j \bs\sigma_j^z \ket{\{p\}}=\ket{\{p\}}$ in Neveu-Schwarz. We have evaluated the expectation value of the flip operator using $\braket{\prod_{j=1}^L\sigma^z_j}=i^L\operatorname{Pf}(\Gamma)$, valid whenever the state is a Slater determinant.

Given the symbol associated with the excited state, we have reconstructed its correlation matrix via a (discrete) Fourier transform. From that we have then computed the half-chain von Neumann entropy using 
$$
S_1(A)=-\mathrm{tr}[\tfrac{\mathrm{I_{2|A|}+\Gamma_A}}{2}\log \tfrac{\mathrm{I_{2|A|}+\Gamma_A}}{2}].
$$
\paragraph*{Simulated annealing.}
Simulated annealing is an algorithm to approximate the global minimum of a function defined in a large search space. Since it does not rely on evaluating any gradient, it is particularly suited to our discrete search space, made by all the eigenstates.
To start off the algorithm we initialise the system in a random eigenstate, obtained by sampling the occupation of each momentum from a uniform distribution. At each step of the algorithm, the state is described by a set of occupied momenta $\{p\}$. The next step is obtained from the previous one 
as follows: a particle-hole transformation at a random momentum is performed, giving a new set of momenta $\{p'\}$; the new state is accepted with probability $\min\{\exp(\frac{S_1(\frac{L}{2};\{p\})-S_1(\frac{L}{2};\{p'\})}{\mathfrak s}, 1\}$ in analogy with the Metropolis–Hastings algorithm.
The reference entropy $\mathfrak s$ we use to rescale the entropies and define the acceptance probability is \emph{not} constant during the algorithm, but depends on the number of iteration: at the beginning we set it large enough not to significantly affect the probability to jump from one state to the other, and it goes to $0$ when approaching the maximum number of iterations. 
In this way, the algorithm initially explores the full landscape, ignoring small features of the entropy, then it drifts towards low-entropy regions and, once one of those regions has been chosen, it finally moves to its local minimum in a (quasi)deterministic way; the state that is finally reached is a candidate for being the global minimum. For the sake of generality, we have run the algorithm several times, also varying how $\mathfrak s$ is updated; in particular, we used power-law decays of the form 
\begin{equation}\label{eq:spar}
    \mathfrak s= \mathfrak s_0\left(\frac{iteration_{tot} - iteration_{current}}{iteration_{tot}}\right)^\alpha,
\end{equation}
where $iteration_{tot}$ is the (chosen) total number of iterations, $iteration_{current}$ is the current iteration and $\alpha$ and $\mathfrak s_0$ are arbitrary positive real numbers.

\subsection{Additional numerical evidence}\label{SM:nummore}

In this section we provide additional numerical data in support of the picture drawn in the main text. 

\begin{figure}
    \centering
    \includegraphics[width=.9\linewidth]{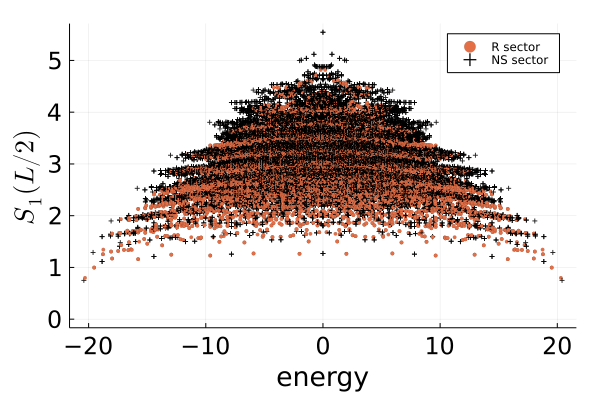}
    \includegraphics[width=.9\linewidth]{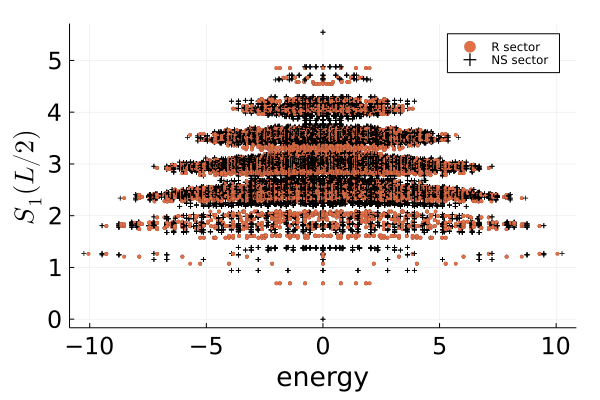}
    \includegraphics[width=.9\linewidth]{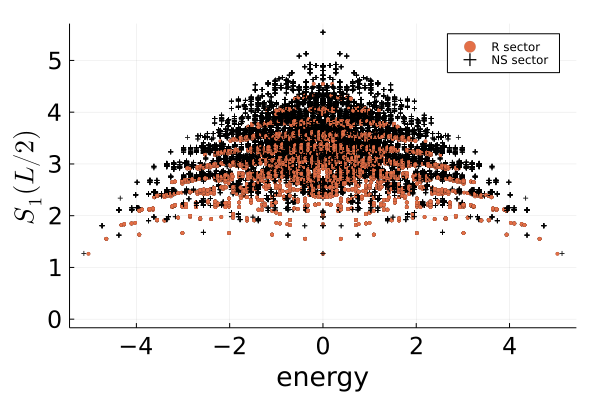}
    \caption{The entropy von Neumann entropy of half chain $S_1(L/2)$ for each eigenstate of a spin chain with $L=16$ in  CI, XX, X-X (from top to bottom). Circles (crosses) correspond to eigenstates belonging to the Ramond (Neveu-Schwartz) sector.}
    \label{fig:entropy_all}
\end{figure}

\begin{figure}
    \centering
    \includegraphics[width=.9\linewidth]{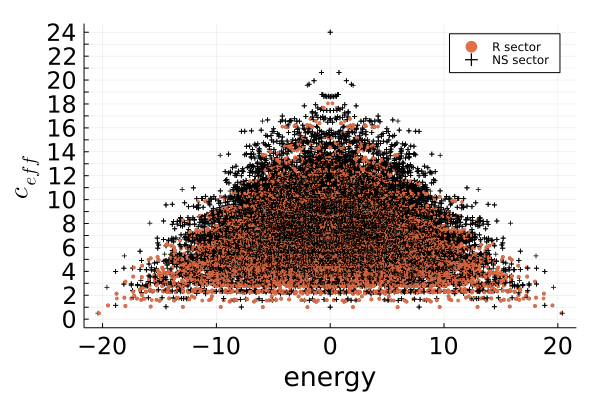}
    \includegraphics[width=.9\linewidth]{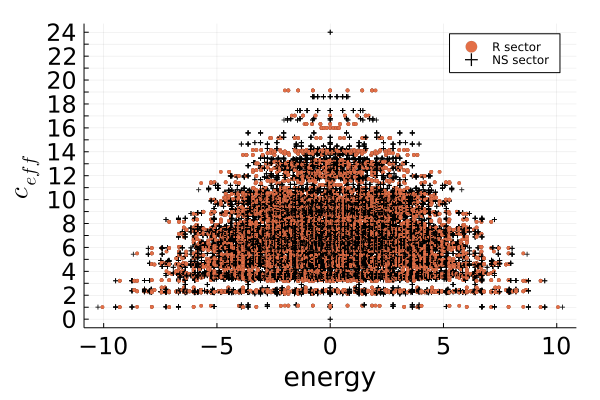}
    \includegraphics[width=.9\linewidth]{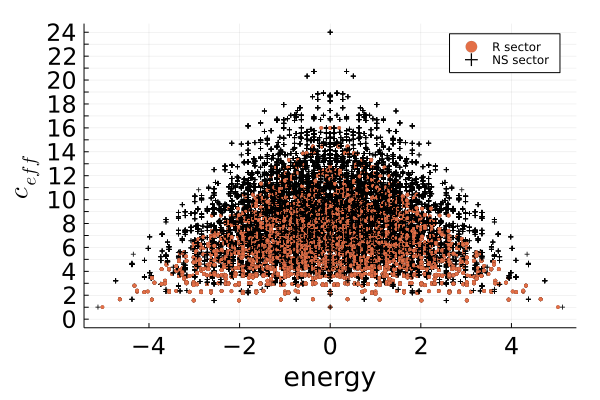}
    \caption{The combination of entropies $c_{eff}\equiv(S_{1}(L/2)-S_{1}(L/4))\frac{6}{\log 2}$ for each eigenstate of a spin chain with $L=16$ in  CI, XX, X-X (from top to bottom). Circles (crosses) correspond to eigenstates belonging to the Ramond (Neveu-Schwartz) sector.}
    \label{fig:entropy_combo_all}
\end{figure}

We recall that our goal is to find the minima of the entropy in the XX, critical Ising, and strongly anisotropic XY model.
The full entropy landscapes in a small chain are reported in Fig~\ref{fig:entropy_all}. In Fig.~\ref{fig:entropy_combo_all} we also show a special combination of entropies $c_{eff}\equiv(S_{1}(L/2)-S_{1}(L/4))\frac{6}{\log 2}$ (analogous combinations have been considered, e.g., in Ref.~\cite{Lauchli_2008}) that remains finite in low-entangled excited states. 
Specifically, it asymptotically gives $0$ in the excited states minimizing gapped operators with quasilocal densities; it instead gives the central charge in the gapless case whenever it can be described by a conformal field theory. The low-entangled excited states that differ from  the aforementioned states  in just a finite number of excitations have generally higher $c_{eff}$, as one can deduce from Refs~\cite{Alcaraz2011Entanglement,Berganza2012Entanglement} for the states with a CFT description. Fig.~\ref{fig:entropy_combo_all} suggests that a similar behaviour should be expected also in the states satisfying the area law --- see also Section~\ref{SM:lower}, indeed we see that in XX only a finite number of excited states exhibit $c_{eff}=0$ whereas all the others have $c_{eff}\gtrsim 1$.

\begin{figure}
    \centering
    \includegraphics[width=.9\linewidth]{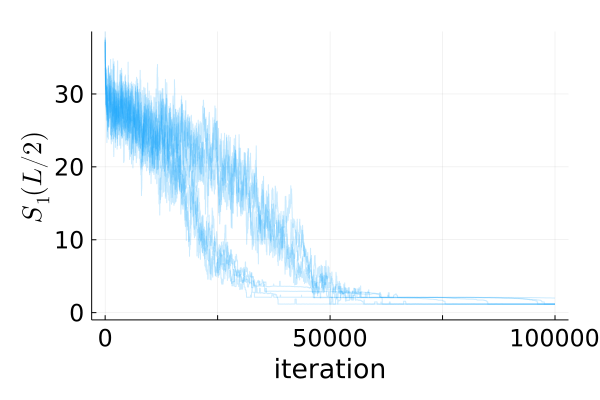}
    \includegraphics[width=.9\linewidth]{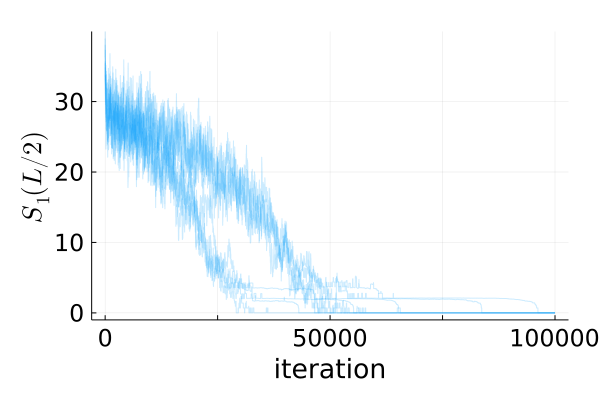}
    \includegraphics[width=.9\linewidth]{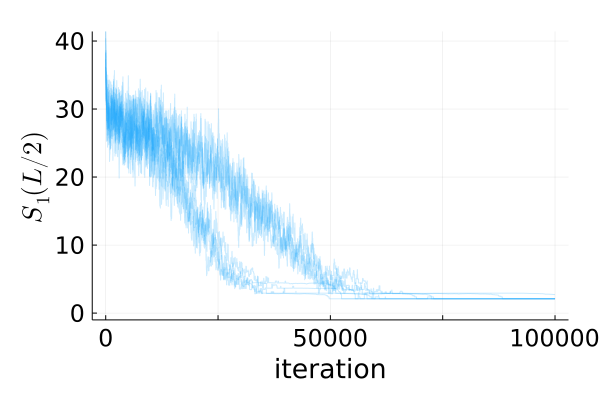}
    \caption{Annealing history of several runs of the simulated annealing algorithm for $L=200$ and $s\in[0.5,1.5]$ --- \eqref{eq:spar}. 
    From top to bottom: CI in Ramond, XX in Neveu-Schwartz, X-X in Neveu-Schwartz.}
    \label{fig:annealing_history_all}
\end{figure}

Fig.~\ref{fig:annealing_history_all} shows the variation of the entropy during several runs of simulated annealing. As the number of iterations increases, the entropy approaches a minimum. The algorithm does not always converge to what we have identified as the absolute minimum: when it does not, we have checked that it gets stuck in a configuration with few domain walls, whose entropy is well described in the thermodynamic limit and grows logarithmically with the system's size.

\begin{figure}
    \centering
    \includegraphics[width=.9\linewidth]{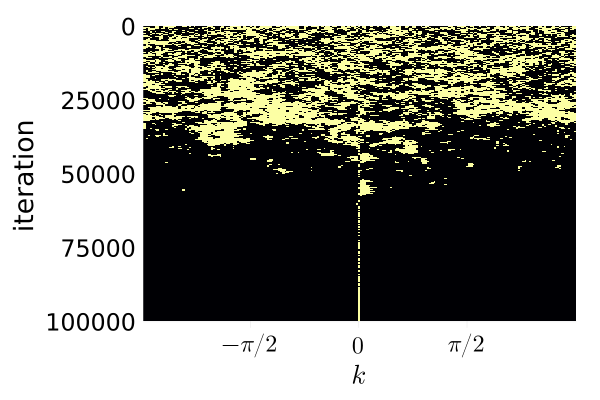}
    \includegraphics[width=.9\linewidth]{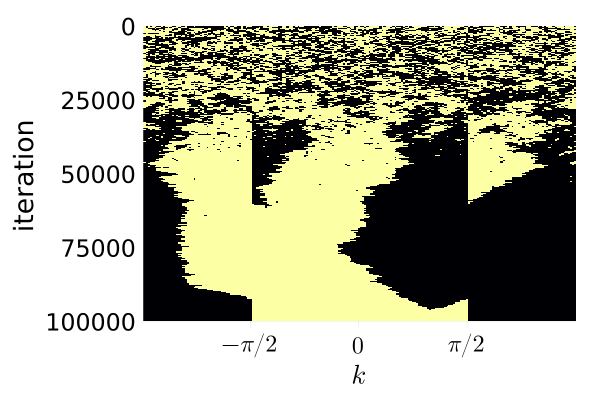}
    \includegraphics[width=.9\linewidth]{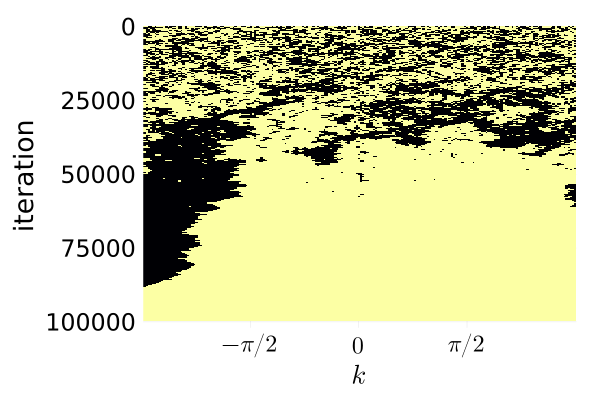}
    \caption{Occupation numbers during a simulated annealing in a chain of length $L=200$.  
    Assuming that the state at a given iteration is described by the set of momenta $\{p\}$, the colour code is: yellow if $k\in\{p\}$, black if $k\notin\{p\}$.
    From top to bottom: CI in Ramond, XX in Neveu-Schwartz, X-X in Neveu-Schwartz.
    }
    \label{fig:percolation_all}
\end{figure}

Fig.~\ref{fig:percolation_all} reports 
the occupation number of all momenta during single runs of simulated annealing.
For the sake of simplicity, we have chosen  the ground state as the vacuum of the excitations. In all the three models the ground state is critical and its entropy grows logarithmically with the system's size.
We see that the occupation numbers are completely uncorrelated as long as $\mathfrak s$ --- \eqref{eq:spar} --- is large enough, then they start forming clusters as $\mathfrak s$ decreases.
In the specific run, the finial state in CI belongs to the ground-state's eigenspace.
The final state in XX is the one with all spins aligned along the $z$ direction: the Fermi sea appearing in the panel exactly compensates the Fermi sea of the ground state. 
The final state in X-X is the maximally excited state.

\subsection{A lower bound for the von Neumann entropy}\label{SM:lower}

\begin{figure}
    \centering
    \includegraphics[width=\linewidth]{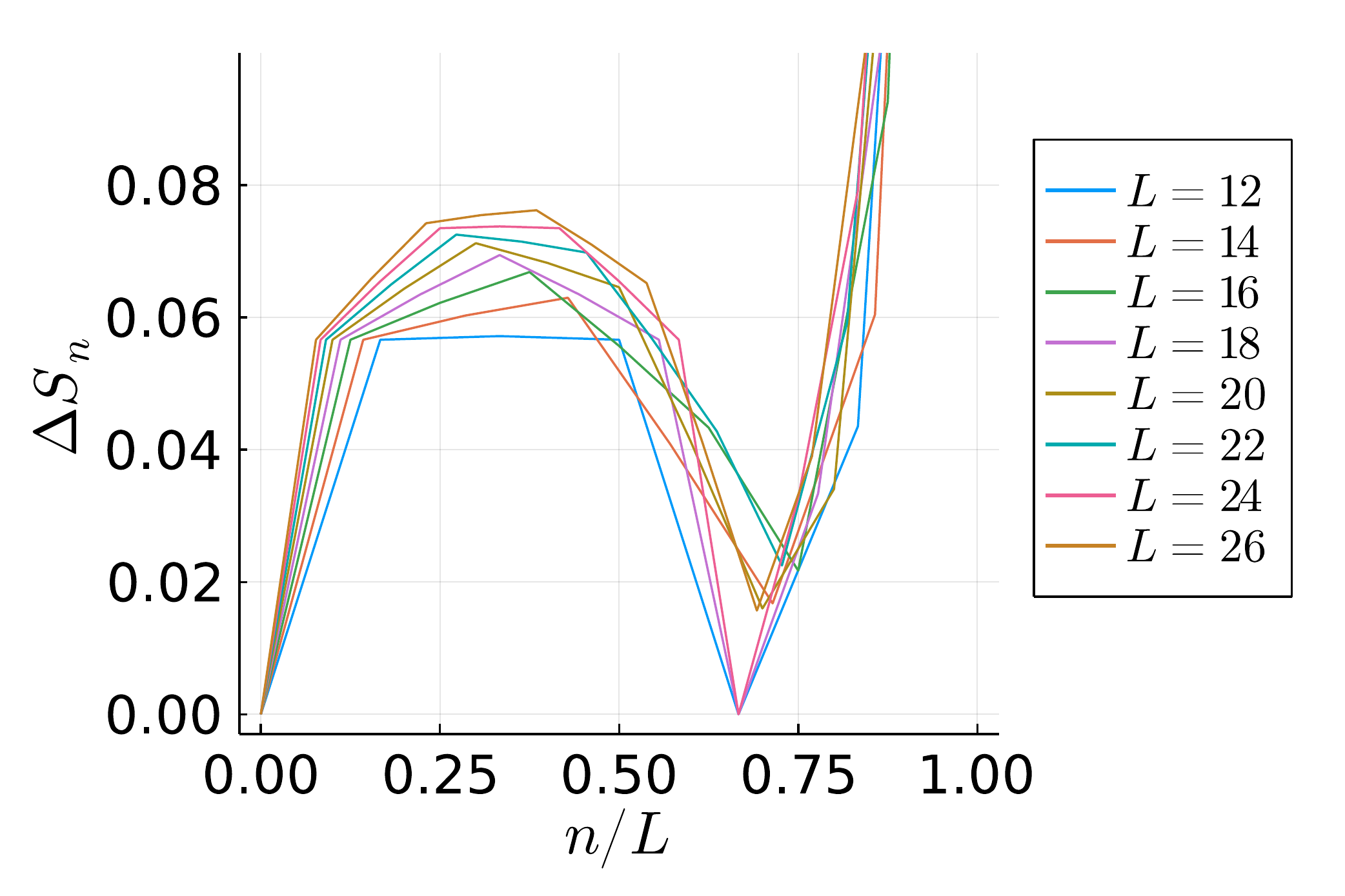}
    \includegraphics[width=\linewidth]{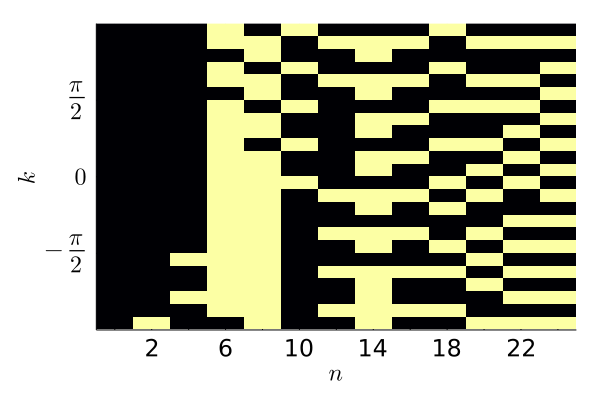}
    \caption{Top: Discrepancy $\Delta S_n$ from the conjectured lower bound of $S_1(L/2)$ in XX for various chain's lengths; here $\Delta S_n=\operatorname{min}(S_1(L/2)|n) - \frac{n}{2}H(1/3)$,
    where $\operatorname{min}(S_1(L/2)|n)$ is the minimum of the entropy among the states with $n$ domain walls with respect to the ferromagnetic eigenstate.
    Bottom: configurations corresponding to $\operatorname{min}(S_1(L/2)|n)$ for $L=24$. In yellow (black) the particle (hole) momenta.}
    \label{fig:minimum_config_xx}
\end{figure}

Ref.~\cite{Alba2009Entanglement} identified a class of excited states with exceptional entanglement properties. Their symbol has a repeated pattern of sign changes and the corresponding entropies exhibit a finite-size profile with a remarkable piece-wise linear behaviour. Here we provide evidence that the states with minimal entropy for given number of changes of sign 
belong to this class. This, in turn, will allow us to conjecture the value of the minimal entropy. 
We start with the XX model and define the excitations with respect to one of the two ferromagnetic excited states. We represent the excited states as domains of occupied neighbourhood momenta alternating with domains of unoccupied ones. 
Our reasoning starts from the following observations:
\begin{itemize}
\item At fixed number of domains, if they have an extensive size, the mapping to a conformal field theory indicates that the half-chain entropy scales as $\frac{c}{3}\log\frac{L}{\pi}$, where the central charge $c$ can be identified with half the number of domains $n$. 
\item The excited states corresponding to scattering the occupied momenta uniformly consist of an effective pattern that repeats over and over again.  There, in the thermodynamic limit ($|A|\ll L$) the von Neumann entropy of a subsystem $A$ scales as $H(\frac{n}{2L})|A|$, where $H(x)=-x\log x-(1-x)\log(1-x)$ is the binary entropy function.
For $n\ll L$ this behaves as $\frac{n|A|}{2L}\log(\frac{2L}{n})$. 
\item If there are domains with different patterns, in the thermodynamic limit the entropy develops a positive logarithmic correction. It is therefore reasonable to expect that the entropy is minimal for the smallest possible number of domains.
\end{itemize}
In view of this, we study the configurations corresponding to a single domain of a repeated pattern in which only one momentum is occupied out of $b$; outside the domain the momenta will be all unoccupied. 
We claim that the von Neumann entropy of a subsystem of length $\ell$ has the following leading asymptotic behaviour:
\begin{equation}
S_1(\ell)\sim 
\tfrac{n}{2}\Bigl[H(\tfrac{2}{b}\lfloor\tfrac{b \ell}{L}\rfloor)+[H(\tfrac{1}{b}\lceil\tfrac{b \ell}{L}\rceil)-H(\tfrac{1}{b}\lfloor\tfrac{b \ell}{L}\rfloor)](\tfrac{b\ell}{L}-\lfloor\tfrac{b \ell}{L}\rfloor)\Bigr]\, .
\end{equation} 
Thus we get
$$
S_1(\tfrac{L}{2})\sim \begin{cases}\frac{n}{2}H(\tfrac{1}{2})& b\text{ even}\\
\frac{n}{2}H(\tfrac{b-1}{2 b})& b\text{ odd.}
\end{cases}
$$
The configuration associated with the minimal half-chain entropy has $b=3$.  
Note, however, that the states we consider exist only if $ n \leq \frac{2L}{b}$ and, for $ n < \frac{2L}{b}$ there is a logarithmic correction $O(\log L)$ due to the partially filled Fermi sea. 
In the specific case minimising the entropy, the expression is valid up to $n\leq \frac{2}{3}L$. For  $n>\frac{2}{3}L$ the states with minimal entropy display more patterns, exhibiting in turn a larger entropy than the naive extension of our estimation $\frac{n}{2}H(1/3)$ outside its range of validity. We therefore conclude
\begin{equation}
S_1(\tfrac{L}{2})\geq \tfrac{n}{2}H(\tfrac{1}{3})\, .
\end{equation}
Fig.~\ref{fig:minimum_config_xx} shows the validity of our conjecture in small chains up to $26$ spins. Note that, for $n=\frac{2L}{3}$ there are no logarithmic corrections and the inequality seems to be exactly saturated.  

\begin{figure}
    \centering
    \includegraphics[width=\linewidth]{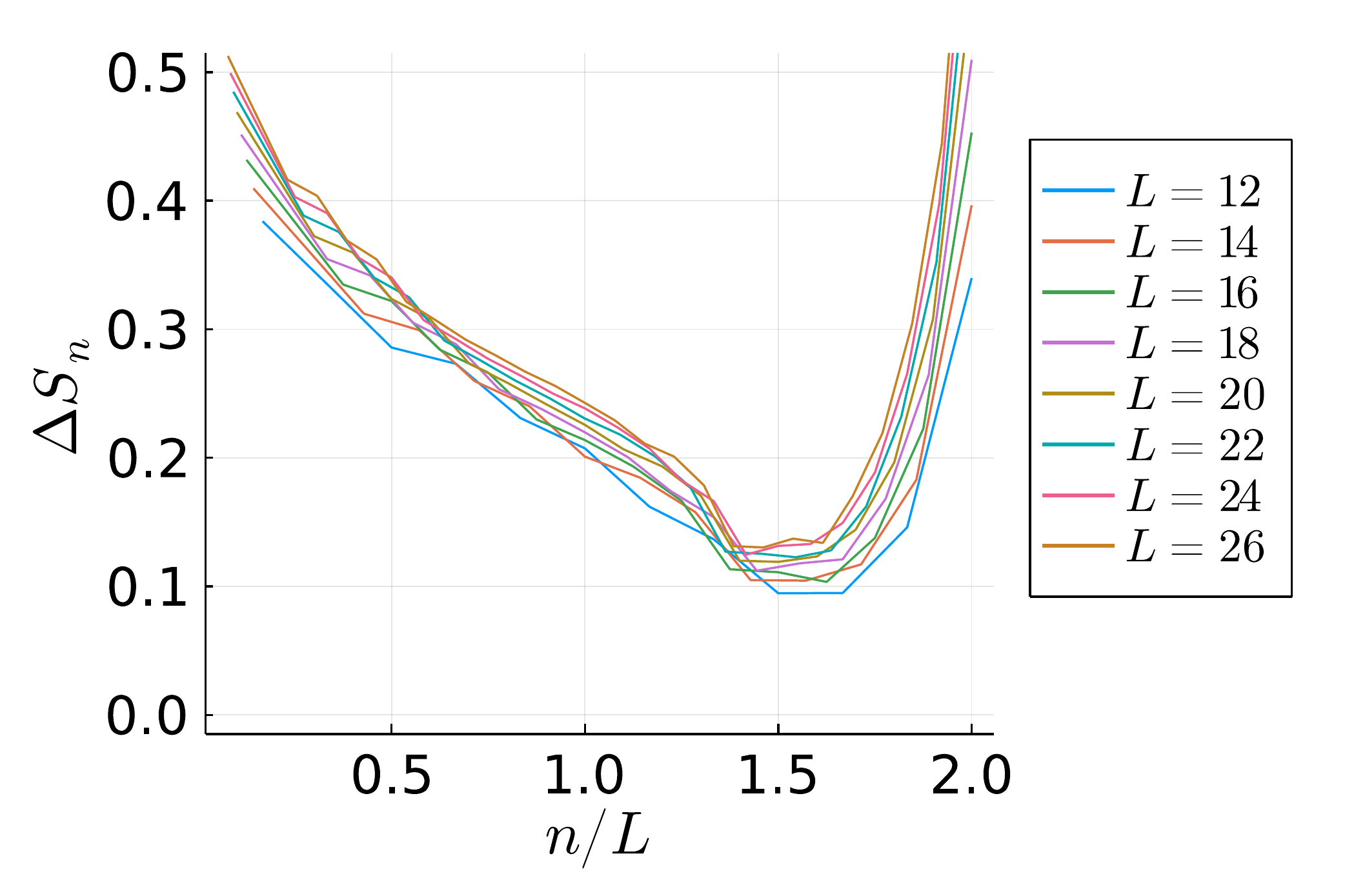}
    \includegraphics[width=\linewidth]{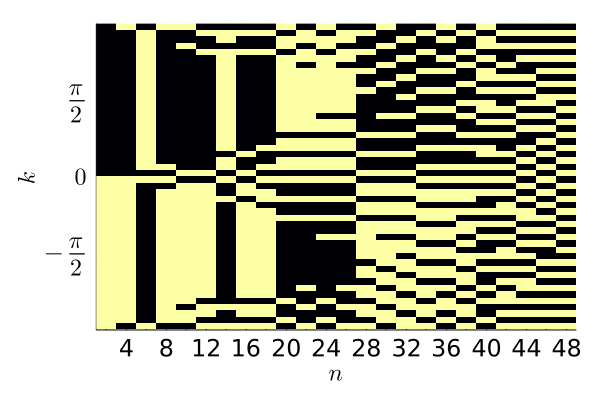}
    \caption{Top: Discrepancy $\Delta S_n$ from the conjectured lower bound of $S_1(L/2)$ in the NS sector of CI for various chain's lengths; here $\Delta S_n=\operatorname{min}(S_1(L/2)|n) - \frac{n}{4}H(1/3)$,
    where $\operatorname{min}(S_1(L/2)|n)$ is the minimum of the entropy among the states with $n$ domain walls in the scalar symbol.
    Bottom: configurations corresponding to $\operatorname{min}(S_1(L/2)|n)$ for $L=24$. In yellow (black) the particle (hole) momenta.}
    \label{fig:minimum_config_ising}
\end{figure}

An analogous (though a little more involved) analysis applies to the critical Ising model if we represent the excited states as domains of pseudo-momenta in which the (antisymmetric!) scalar symbol is $1$ alternating with domains in which it is $-1$. We then find the bound
\begin{equation}
S_1(\tfrac{L}{2})\geq \tfrac{n}{4}H(\tfrac{1}{3})\, .
\end{equation}
The validity of this conjecture is shown in Fig.~\ref{fig:minimum_config_ising}.

\subsection{On the equivalence of models}\label{SM:equivalence}
We use here the label `one-site shift invariant model' to indicate the algebra of the one-site shift invariant local conservation laws. We then say that two models are equivalent if they can be mapped to one another by a unitary transformation that preserves the locality properties of either the spins or the underlying fermions. As far as the entanglement of a connected block of spins is concerned, the fermionic bipartite entanglement is equivalent to the spin one~\cite{Vidal2003Entanglement,Fagotti2010disjoint}, hence we can allow transformations preserving the locality of either fermions or spins. 
We identify such pseudolocal classes of equivalence within noninteracting spin chains and exhibit what is arguably the simplest representative of each class. 
We consider one-site shift invariant noninteracting Hamiltonians with short-range or exponentially fast decaying interactions.
Their most general 2-by-2 symbol reads
\begin{equation}
\hat{\mathcal H}^{(2)}(k)=w(k)\mathrm I+\vec \varepsilon(k)\cdot \vec \sigma
\end{equation}
where $w(-k)=-w(k)$ and $\varepsilon_j(-k)=-(-1)^j \varepsilon_j(k)$ are smooth functions of $k$. 
Since any operator with symbol proportional to the identity (such as the quadratic Dzyaloshinskii-Moriya interaction) commutes with the Hamiltonian, without loss of generality we can set $w(k)=0$ (it does not affect the set of one-site shift invariant charges). For the sake of simplicity we also assume $\varepsilon_3(k)=0$; the generality of this assumption will be addressed later. We start considering a one-site shift invariant transformation generated by an operator with exponentially decaying interaction. By applying such a quasilocal transformation to the XX model, we end up with a model characterised by a symbol of the form
\begin{multline}\label{eq:mapping1}
\varepsilon(k)e^{i\frac{\theta(k)}{2}\sigma^z}\sigma^y e^{-i\frac{\theta(k)}{2}\sigma^z}=\\
\varepsilon(k)\cos\theta(k)\sigma^y+\varepsilon(k)\sin\theta(k)\sigma^x.
\end{multline}
If this symbol is smooth, it parametrises a generic model of the kind discussed above. For it to be smooth
it would be sufficient that $\varepsilon^2(k)$ is smooth and $\theta(k)$ is smooth up to odd jumps multiple of $\pi$, which could be compensated by changes of sign in $\varepsilon(k)$. Our locality assumptions on the transformation, on the other hand, require $\theta(k)$ to be an odd smooth function in $(-\pi,\pi)$ extended by periodicity. 
As a matter of fact, we can always redefine $\theta(k)$ so as to remove the jumps in $]-\pi,0[\cup ]0,\pi[$ at the price of potential jumps at $k=0$ and $k=\pm\pi$. We can also shrink the discontinuity at $k=0$ as much as possible, which means that it will become either $0$ (hence, $\theta(k)$ continuous in its neighbourhoods) or $\pi$. The discontinuity at $k=\pi$ will then be any multiple of $\pi$. After these redefinitions, the equivalence between the generic model represented by the right hand side of \eqref{eq:mapping1} and the XX model  is established only when $\theta(k)$ is smooth at $k=0$ and $k=\pi$. But this is not the end of the story. 
Let us  consider indeed the case in which $\theta(k)$ is continuous at $k=0$ but has a $2n\pi$ discontinuity at $k=\pm \pi$. 
We can write
$$
\theta(k)=\theta_0(k)+n k\quad k\in(-\pi,\pi)\, ,
$$
where $\theta_0(k)$ is smooth. Since the discontinuity at $k=\pm \pi$ does not change the sign of $\sin\theta(k)$ and $\cos\theta(k)$, $\varepsilon(k)$ can be any smooth even function of $k$. This means that a transformation preserving the spin locality maps the model into one characterised by the symbol
$$
\epsilon(k)\sigma^y e^{-i n k \sigma^z}\, ,
$$
with $\epsilon(-k)=\epsilon(k)$ a smooth function of $k$.
Let us now consider local unitary transformations that are not generated by operators with exponentially decaying interactions. A relevant example 
is the Kramers-Wannier duality corresponding to shifting the Majorana fermions $\bs a_j$ by one pseudo-site. It is easy to show that it is translated into the following mapping
$$
\epsilon(k)\sigma^y e^{-in k \sigma^z}\mapsto -\epsilon(k) \sigma^y e^{i(n-1) k \sigma^z}\, .
$$
This allows us to restrict to even $n=2m$, i.e., symbols of the form
$
\epsilon(k) \sigma^y e^{-2i m k \sigma^z}
$.
Remarkably, we can also get rid of $e^{-2i m k \sigma^z}$ by shifting the Majorana fermions by $m$ pseudosites in opposite directions depending on the parity of their pseudosite. 
And the model is finally mapped into  XX
$$
\epsilon(k)\sigma^y e^{-i n k \sigma^z}\mapsto \epsilon(k)\sigma^y\, .
$$

Let us now consider the case in which  $\theta(k)$ has a $\pi$ discontinuity at $k=0$ and a $(2n+1)\pi$  discontinuity at $k=\pi$.  We can write
$$
\theta(k)=\theta_0(k)+\tfrac{\pi}{2}\mathrm{sgn}(\sin(k))+n k\, ,
$$
and the model can be mapped into
$$
\epsilon(k)\sin k\sigma^xe^{-i n k \sigma^z}
$$
A Kramers-Wannier duality is translated into the mapping
$$
\epsilon(k)\sin k\sigma^xe^{-i n k \sigma^z}\mapsto \epsilon(k)\sin k\sigma^xe^{i (n-1) k \sigma^z}
$$
which allows us again to focus on even $n=2m$ and apply the same asymmetric shift as before to the Majorana fermions. This results in the symbol
$$
\epsilon(k)\sin k\sigma^x\, .
$$
Interestingly, for $\epsilon(k)=1$ the associated Hamiltonian can be mapped to the XX one by the spin flip $\prod_j \bs\sigma_{2j-1}^y\bs\sigma_{2j}^z$, which however breaks one-site shift invariance. One could then map the one-site shift invariant excited states of this models to a basis of excited states of the XX model that generally breaks one-site shift invariance.

Let us now consider $\theta(k)$ continuous around $k=0$ and discontinuous with a $(2n-1)\pi$ discontinuity at $\pi$.  We can write
$$
\theta(k)=\theta_0(k)+(n-\tfrac{1}{2})k\, .
$$
While $\sin\theta(k)$ and $\cos\theta(k)$ do not change sign at $k=0$, they do it at $k=\pi$, which should be compensated by a change of sign of $\varepsilon(k)$. Since $\varepsilon(k)$ can not be discontinuous at $k=\pi$, it has to vanish. A transformation generated by a quasilocal operator maps therefore the model into
$$
\epsilon(k)\cos \tfrac{k}{2} \sigma^y e^{-i (n-\frac{1}{2}) k \sigma^z}\, .
$$
A Kramers-Wannier duality is translated into the mapping
$$
\epsilon(k)\cos \tfrac{k}{2} \sigma^y e^{-i (n-\frac{1}{2}) k \sigma^z}\mapsto -\epsilon(k)\cos \tfrac{k}{2} \sigma^y e^{i (n-\frac{3}{2}) k \sigma^z}\, ,
$$
which allows us again to restrict ourselves to even $n=2m$ and apply the same asymmetric shift as before to the Majorana fermions. This results in the symbol
$$
\cos \tfrac{k}{2} \sigma^y e^{-i \frac{k}{2} \sigma^z}\, ,
$$
which represents the critical Ising model with negative magnetic field. Note that the sign of the field can be changed by a spin flip in the $x$ direction, which is not captured by the transformations considered so far because its symbol does not have a 2-by-2 representation.  

The last case to consider is when 
there is a $\pi$ discontinuity at $k=0$ and a $2n \pi$ discontinuity at $k=\pi$, that is to say,
$$
\theta(k)=\theta_0(k)+\tfrac{\pi}{2}\mathrm{sgn}(\sin(\tfrac{k}{2}))+(n-\tfrac{1}{2})k\, ,
$$
with $\theta_0(k)$ smooth. While $\sin\theta(k)$ and $\cos\theta(k)$ do not change sign at $k=\pi$, they do it at $k=0$, which should be compensated by $\varepsilon(k)$. Since $\varepsilon(k)$ can not be discontinuous at $k=0$, it has to vanish. The model can then be mapped into
$$
\epsilon(k)\sin \tfrac{k}{2} \sigma^x e^{-i (n-\frac{1}{2}) k \sigma^z}
$$
Again, a Kramers-Wannier duality is translated into the mapping
$$
\epsilon(k)\sin k\sigma^xe^{-i (n-\frac{1}{2}) k \sigma^z}\mapsto \epsilon(k)\sin k\sigma^xe^{i (n-\frac{3}{2}) k \sigma^z}\, ,
$$
which allows us to assume $n$ even and apply the asymmetric shift as before to the Majorana fermions, finally bringing the symbol to
$$
\epsilon(k)\sin k\sigma^xe^{-i \frac{k}{2} \sigma^z}\, ,
$$
which corresponds to the critical Ising model (with positive magnetic field). 

We now address the problem of generality of the symbol we started from, i.e., the fact that we assumed $\varepsilon_3(k)=0$. In particular, we wonder whether a more generic symbol can always be mapped into the one we considered. We parametrise the most generic symbol as follows
\begin{multline}
e^{i\frac{\phi(k)}{2}\sigma^y}\mathcal H^{(2)}(k) e^{-i\frac{\phi(k)}{2}\sigma^y}=\\
\cos \phi(k)\mathcal H^{(2)}(k)+\sin \phi(k) \varepsilon(k)\sin \theta(k)\sigma^z\, ,
\end{multline}
where $\mathcal H^{(2)}(k)$ belongs to the family of symbols considered above. Analogously to the previous case, the smoothness of the right hand side of the equation translates into $\phi(k)$ being smooth up to $\pi$ discontinuities, which can again be fixed by a proper choice of the sign of $\varepsilon(k)$. Since $\phi(k)$ is even by the symmetry constraints on the symbol, all the discontinuities are removable and hence the operator with symbol $\frac{\phi(k)}{2}\sigma^y$, which generates the desired transformation, can always be chosen quasilocal. This shows that, without loss of generality, we can impose $\varepsilon_3(k)=0$.

\end{document}